\documentclass[preprint,12pt]{elsarticle}

 \usepackage{graphicx}

\usepackage{amssymb}
\usepackage{amsthm}
\usepackage[nodots]{numcompress}

\journal{Materials Letter}

\begin{document}

\begin{frontmatter}

\title{The Crystallization Behavior of Porous PLA Prepared by Modified
Solvent Casting/Particulate Leaching Technique for Potential Use of Tissue
Engineering Scaffold}

\author[sjtu]{Ran ˜Huang \corref{cor1}}
\ead{ranhuang@sjtu.edu.cn}
\author[sjtu]{Xiaomin ˜Zhu}
\ead{zxmin0608@sjtu.edu.cn}
\author[sjtu]{Haiyan ˜Tu}
\ead{494628103@qq.com}
\author[sjtu,tju]{Ajun ˜Wan \corref{cor2}}
\ead{wanajun@sjtu.edu.cn}
\cortext[cor1]{Corresponding author}
\cortext[cor2]{Principal corresponding author}
\address[sjtu]{School of Chemistry and Chemical Engineering, Shanghai JiaoTong University, Shanghai 200240, China}
\address[tju]{State Key Laboratory of Pollution Control and Resources Reuse, National Engineering Research Center of Facilities Agriculture, Tongji University, Shanghai 200092, China}

\begin{abstract}
The porous PLA foams potential for tissue engineering usage are prepared by
a modified solvent casting/particulate leaching method with different
crystallinity. Since in typical method the porogens are solved in the
solution and flow with the polymers during the casting and the crystallinity
behavior of PLA chains in the limited space cannot be tracked, in this work
the processing is modified by diffusing the PLA solution into a steady salt
stack. With a thermal treatment before leaching while maintaining the stable
structure of the porogens stack, the crystallinity of porous foams is made
possible to control. The characterizations indicate the crystallization of
porous foams is in a manner of lower crystallibility than the bulk
materials. Pores and caves of around 250${%
\mu%
m}$ size are obtained in samples with different crystallinity. The
macro-structures are not much impaired by the crystallization nevertheless
the morphological effect of the heating process is still obvious.
\end{abstract}

\begin{keyword}
Porous materials \sep Biomaterials \sep Poly(Lactic Acid) \sep Tissue Engineering Scaffold \sep Crystallization
\end{keyword}

\end{frontmatter}

\section{INTRODUCTION}

Porous poly(lactic acid) (PLA) has been developed for tissue engineering
scaffolds for decades [1-3]. Including poly(L-lactic acid) (PLLA),
poly(D-lactic acid) (PDLA) and PLA-based copolymers like
poly(lactic-co-glycolic) acid (PLGA), these bio-based resins have been
proved to be a successful candidate of scaffold materials with excellent
biocompatibility and biodegradablity. The tissue engineering requires
sufficient interconnecting inner space in the scaffold for biofactor
delivery, tissue growth, and the scaffold should be degradable after
tissue's growth meanwhile providing proper mechanical strength to support
the tissue engineering system. Therefore the balance of growth space,
degradation behavior and mechanical properties is the main concern of
constructing a scaffold. With the respect to the particular requirements of
certain tissue engineering, nowadays designed preparation and modification
techniques of porous PLA scaffold materials become an intensive
interests-drawing subject [4-13], which requires more understanding of the
basic principles of physical and chemical properties, particular in the form
of scaffold.

The crystallization of PLA plays an important role in its mechanical
properties and degradability. Generally the crystallized polymers have
higher strength and mechanical modulus [14]. In the case of PLA, the
crystallinity also significantly affects the degradability, with the general
behavior that the degradation time is longer with higher crystallinity, as
the crystal segments are more stable than amorphous area and prevent water
permeation into it. For example, it was reported that PLLA takes more than 5
years for total degradation, whereas only about 1 year for the amorphous PLA
or PDLLA [15]. However unlike the crystallinity control for the inorganic
components in the tissue engineering scaffold [16], very rare reports
concerned the crystallinity of the polymer scaffold materials. One probable
reason is that the preparation of porous scaffolds is a delicately process,
where the control of crystallinity is usually difficult or unavailable.
Except the solvent casting/particle leaching method, in other widely used
preparation methods such as electrospun fiber, phase separation, membrane
lamination and gas foaming, the polymers are not able to experience a
thermal treating step, i.e. the most common way to control the crystallinity
[1]. In some works the crystallinity is controlled by the raw materials
itself, i.e. selecting raw materials of different molecular weight
associated with different crystallization behaviors, or a particular
processing procedure for chain cleavage to control the crystallinity [17].
And in some methods, even polymer with high crystallinity cannot be served
as the raw materials to prepare the scaffold, for example the gas foaming
technique reported by David J. Mooney et al. [18]. Nevertheless, the study
of the crystallization of the scaffold should hold considerable practical
merits in tissue engineering, for example to fit the scaffold degradation
time with the expected tissue growth time by the control of crystallinity
(if possible). Also, the scaffold structure may vary with crystallinity and
influent the biological behavior of living tissue leaning on it. Park et al.
reported a research on the sustained release of human growth hormone from
semi-crystalline poly(l-lactic acid) and amorphous
poly(d,l-lactic-co-glycolic acid) microspheres, which reveals that the
morphological effect is important on protein release [19].

Crystallinity can be tailored in solvent casting/particulate leaching
technique. However in this method the salt as porogens are solved in the
solution and flow with the polymers during the casting, therefore without
the immobilization of porogens the crystallinity behavior in the
space-limited gap cannot be tracked [1]. In this work, we modified the
solvent casting/particulate leaching technique by diffusing the PLA solution
into a steady salt stack instead of solving the porogens. The control of
crystallinity was made possible by inserting a thermal treating step before
leaching, while maintaining the stable structure of salt stack. We have
investigated the morphological effect of limited space on the
crystallization of PLA, and the porous structure with different
crystallinity under thermal treatments.

\section{Experimental}

\subsection{Materials}

The PLA of label 4032D is purchased from Natureworks\circledR , with L/D
ratios from $24$:$1$ to $30$:$1$. The porogens is NaCl of analytical grade.
The $1$:$1$ mixture of dichloromethane and chloroform is served as solvent.

\subsection{Porous sponge preparation}

The PLA pellets are solved into dichloromethane and chloroform ($1$:$1$)
with the concentration of $0.2{g}/{l}$. The NaCl powder is
thoroughly grounded and sieved with $109{%
\mu%
m}$ then $300{%
\mu%
m}$ sieve to screen the particles of sizes in between, and paved onto a
petri dish where it forms a \symbol{126}$1.5{mm}$ thick disc. The PLA
solution is very slowly poured into the dish at the edge. The pouring is as
slow as that the solution diffuses inside the salt stack instead of flowing
over the surface, also the slow diffusing guarantees the salt particles are
not considerably moved by the liquid flowing to keep the inner structure of
the salt stack stable. After pouring, the salt-PLA solution composite is
then placed in vacuum for $12{h}$ for drying out the solvent. The
product is a solid dry PLA-glued salt composite ready for thermal treatment.
The composite is placed in water for $48{h}$ to leach out the salts
after the thermal treatment for recrystallization. The leached samples are
freeze-dried for $2{h}$ and stored in vacuum ready for
characterizations.

\subsection{Recrystallization}

The recrystallization of composite is processed by the heating and cooling
process. Four samples were made to have different crystallinity. One sample
was kept as the original composite without thermal treatment for reference
(sample R). The other three composites are heated in oven at $165{%
{}^{\circ}{\rm C}%
}$ for $0.5{h}$, then one composite (sample A) was immediately quenched
in liquid nitrogen; sample B was linearly cooled down at the rate $10{%
{}^{\circ}{\rm C}%
}/30$ ${min}$.; sample C was linearly cooled at $10{%
{}^{\circ}{\rm C}%
}/30$ ${min}$ till $105{%
{}^{\circ}{\rm C}%
}$, kept at $105{%
{}^{\circ}{\rm C}%
}$ for $24{h}$, then cooled down with the same rate to room
temperature. For comparison we also made two bulk PLA samples with the same
crystallization process of sample B and C and they were labeled as sample B'
and C'.

\subsection{Characterization}

The crystallinity of samples were characterized by X-ray Diffraction (XRD)
(D/max-2200/PC, Rigaku Corporation). The porous structure of samples were
revealed by Scanning Electron Microscope (SEM) (Nova NanoSEM 450, FEI).

\section{Results and Discussion}

\subsection{The crystallization behavior}

The XRD results firstly dispel the doubt about the possibility of
crystallization in confined geometry. Fig.1(1) shows the XRD of
three thermal treated samples A, B and C (Note that the sample R is not
included because its XRD curve almost overlap with the sample A, i.e.
amorphous), and the bulk samples of B' and C' are shown in Fig.1(2). With the comparison of the bulk behavior, clear crystal peaks at $%
2\theta =16.6{%
{{}^\circ}%
}$ and other minor peaks indicate that the heat treatment makes the
crystallization possible but we can see significant impact of confined
geometry, the crystallization of PLA chains in porogens slits is harder and
the crystallinity is lower than the bulk sample under the same treatment
condition. The crystallinity is calculated to be $17.11\%$ and $17.43\%$ in
B and C, comparing to $23.86\%$ and $25.35\%$ in B' and C'. It is clear that
with the same thermal treatment the bulk samples are much easier to
crystallize with higher crystallinity. This agrees with our expectation that
the crystallization is more difficult in confined geometry, even the size of
fibers and walls confining the pores is still in the magnitude of
micrometers (see the SEM paragraph) which far overweighted the diameter of
chain segments, the chain movement and rearrangement is obstructed by the
limited space.

Another evidence that confined space impairs the crystallibility is the
comparable crystallinity of B and C. The small difference is trivial due to
many factors such as sample preparation or the baseline selection. It
implies that $7{h}$\ linear cooling process may reach the upper limit
crystallinity for the porogens confined environment, while for bulk sample,
the effects of staying at $105{%
{}^{\circ}{\rm C}%
}$\ for longer time is significant on crystallinity.

Although the samples are leached for $48{h}$ there are unavoidable
porogens residues left in the sample. The peaks on $2\theta =31.7{%
{{}^\circ}%
}$ and $45.5{%
{{}^\circ}%
}$ are NaCl crystals [20], whereas these two peaks do not exist in the bulk
samples B' and C'.

\begin{figure}

    \centering{
	\includegraphics[width=0.8\textwidth]{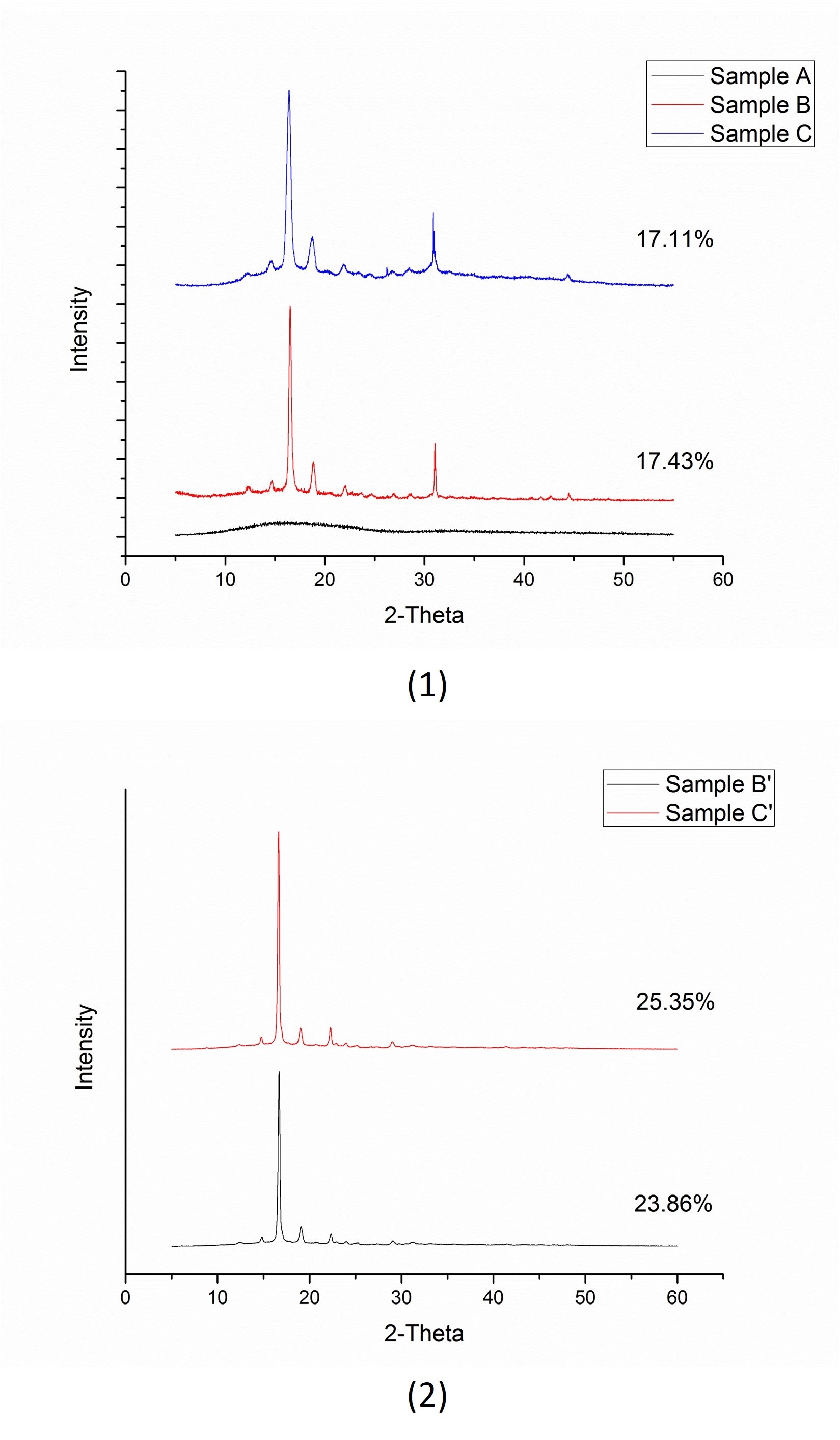}
	\caption{The XRD characterizations of (1): heat-treated porous samples A, B and C, and (2): Bulk samples B' and C' with the same heat treatment.}
	\label{fig1}
    }
\end{figure}
\begin{figure}
    \centering{
	\includegraphics[width=\textwidth]{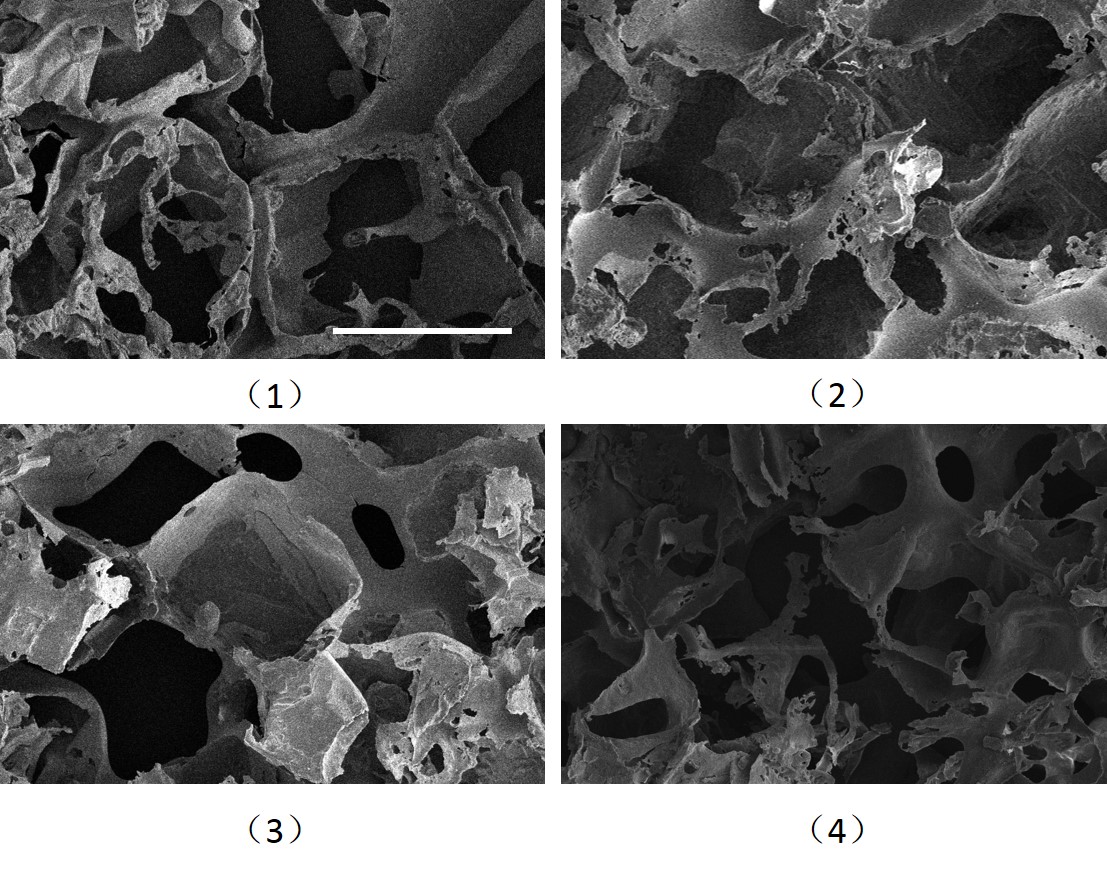}
	\caption{The SEM pictures of four porous
	samples: (1) The reference sample R without heat treatment; (2) Sample A
	with quenching; (3) Sample B with linear cooling; (4) Sample C with linear
	cooling besides being kept at 105${%
	{}^{\circ}{\rm C}%
	}$ for 24 hours. The scale bar in the first graph is 250${%
	\protect\mu%
	m}$. All the photos have the same scale magnitude.}
	\label{fig2}
    }
\end{figure}

\subsection{\protect\bigskip The porous macro-structure}

The SEM results in Fig.2 show the macro-structures of unheated
original casting sample R and three samples with heat treatment A, B and C.
A general observation of macro-structure confirms that the thermal treatment
does not impressively affect the porous structure forming. The pores and
caves structure in each sample can be clearly observed with the pore size of
around $250{%
\mu%
m}$, which accords to the $109\symbol{126}300{%
\mu%
m}$ sieving process. Nevertheless the morphological effect of heat treatment
is obvious. In Fig.2(2), (3) and (4) the reheated samples present
the features of thinner pore walls, branches and fragments, while in the
reference sample the pore wall is thicker with rod-like branches. Regardless
of the crystallinity, heating the samples to $165{%
{}^{\circ}{\rm C}%
}$ (the melting temperature of PLA) enables the polymer chains to remobilize
and diffuse into the slits between salt particles where the solved chains
had not diffused into and occupied. The thinner pore wall indicates that the
chain remobilization also moves the porogens and makes narrower space among
them. Although we employ the steady salt stack to confine the
recrystallization within the limited space, the porogens are only relatively
"stable" comparing to typical solvent casting technique.

The quenching sample A has more fragmental structures than B and C, it is
clear to understand the phenomenon that in sample A the polymer melt
diffused into thinner slits is quenched to solidify its diffusing state of
fragmental features. For sample B and C, the recrystallization process
offers sufficient time for the diffused chains to mobilize and rearrange
themselves to be more ordered, crystal structure. This rearrangement
provides a less fragmental structure on the macro-scale. No obvious
difference of macro-structures between sample B and C is observed, i.e. $7$
hours linear cooling is sufficient for recrystallization to achieve this
structural effect, longer recrystallization time plays very little more
effects on the crystallinity and the macro-structure, and this observation
also agrees to the crystallinity results of B and C as indicated above.

\section{Conclusion}

The PLA porous matrix for potential use in tissue engineering have been
prepared by modified salt casting and particulate leaching technique. The
PLA solution is diffused into relative stable salt stack, instead of solving
salts with polymers in typical method. Because the raw salt casting
solidifies the salt-PLA composite, we are able to insert a step of thermal
treatment to recrystallize the polymer matrix before leaching process. In
this way we are able to: 1) investigate the crystallization behavior of PLA
confined in limited space; 2) develop an available crystallinity control
option in porous PLA scaffold preparation. The XRD results indicate the
crystallization of porous foams, in a manner of lower crystallibility than
the bulk materials. The marco-structure of porous samples are observed by
SEM, by obtaining the pores of around $250{%
\mu%
m}$, it is revealed that the polymer foam may crystallize without
significant structure damage. The features of thinner pore walls, branches
and fragments confirmed the effect of heating treatment. Both XRD and SEM
results of sample B and C indicate that 7 hours linear cooling is sufficient
to achieve certain crystallinity and marco-structure.



\end{document}